# Reduction of Thermal Conductivity by Nanoscale 3D Phononic Crystal

Lina Yang[1], Nuo Yang[2a], Baowen Li[1,2b]

[1]Department of Physics and Centre for Computational Science and Engineering, National University of Singapore, Singapore 117542, Republic of Singapore

[2]NUS-Tongji Center for Phononics and Thermal Energy Science, Department of Physics, Tongji University, 200092 Shanghai, China

*To whom correspondence should be addressed.

E-mail: a) imyangnuo@tongji.edu.cn (NY); b) phononics@tongji.edu.cn (BL)

**Abstract**

The thermal conductivity of nanostructures needs to be as small as possible so that it will have a greater efficiency for solid-state electricity generation/refrigeration by thermoelectrics. We studied how the period length and the mass ratio affect the thermal conductivity of isotopic nanoscale 3D phononic crystal of Si. Simulation results by equilibrium molecular dynamics show isotopic nanoscale 3D phononic crystal can make a significance reduction on the thermal conductivity of bulk Si at high temperature (1000 K). Size and mass effects are obvious in manipulating thermal conductivity. The thermal conductivity decreases as the period length and mass ratio increases. The phonon dispersion curves show the decrease of group velocities in 3D phononic crystals. The phonon's localization and band gap is clearly shown in spectra of normalized inverse participation ratio in nanoscale 3D phononic crystal structure.

**Introduction**

Thermoelectric materials are important for generating electricity from waste heat and being used as solid-state Peltier coolers. Improving the efficiency of thermoelectric materials are critical for their applications.[1] The performance of thermoelectric materials depends on the figure of merit ZT,[2] $ZT = S^2\sigma T/\kappa$, where S, T, $\sigma$, and $\kappa$ are the Seebeck coefficient, absolute temperature, electrical conductivity and total thermal conductivity, respectively. ZT can be increased by increasing S or $\sigma$, or decreasing $\kappa$. However, it is difficult to improve ZT in conventional materials. First, simple increase S for general materials will lead to a simultaneous decrease in $\sigma$.[2,3] Also, an increase in $\sigma$ leads to a comparable increase in the electronic contribution to $\kappa$.[2,3] An alternative way to increase ZT is to reduce the thermal conductivity without affecting electronic property.[2,4] Moreover, ultra-low thermal conductivity is also required to prevent the back-flow of heat from hot end to cool end. Therefore, reduction of thermal conductivity is crucial in thermoelectric application.

Phononic crystals are constructed by a period array of scattering inclusions distributed in a host material. Due to its period change of the density and /or elastic constants, phononic crystals exhibit phononic band gaps.[5] This remarkable property is very different from those of the conventional materials and can be engineered to achieve new functionalities. It is demonstrated that superlattice crystals, one dimensional periodic arrangement of two materials, are effective to achieve very low thermal conductivity.[6-10] Superlattices are constructed by period layers of different materials and have quite low thermal conductivity.[11,12] Superlattices have been extensively studied to design thermoelectric materials with high ZT. Preliminary works show there is a minimum value of thermal conductivity in the direction perpendicular to the planes of superlattice when the period length is reduced to nanoscale.[13-15] It is predicted that atomic-scale 3D phononic crystal of Ge quantum-dot in Si has very low thermal conductivity in all three spatial directions.[16] The thermal conductivity is reduced by several orders of magnitude compared with bulk Si. This reduction of thermal conductivity is due to the reduction in group velocities and multiple scattering of particle-like phonons.

Recently, it is demonstrated experimentally that the Si nanomesh film exhibited low thermal conductivity[17] by modification of phonon band structure. Single crystalline Si by phononic crystal patterning in 2D has a smaller value of thermal conductivity (~2 W/m-K) than bulk Si because of the low group velocities and the coherent phononic effects. [18]

In this letter, we study the thermal conductivity of nanoscale 3D silicon phononic crystal. The 3D crystal consists of $^{28}$Si atoms and "isotopes" $^{M}$Si atoms which have the same properties as [28]Si except the mass, where M is the atomic mass of the isotope of Si. The mass ratio, R, is defined as R=M/28. The 3D phononic crystal could also be called 3D superlattice because different material arranged periodically in three spatial directions. We find the 3D phononic crystal has the ability to flatten phonon dispersion compared with that of bulk Si and it could show band gaps when properly arranged. We studied how the period length and the mass ratio affect the thermal conductivity of the 3D phononic crystal. The phonon dispersion curves and inverse participation ratio of the 3D phononic crystal are also computed to understand the mechanism of the reduction of thermal conductivity. On the other hand, the scatterings of isotopic doping could reduce the lattice thermal conductivity significantly without much reduction of the electrical conductivity.[2,15,19] Then, the value of ZT could be improved largely in isotopic nanoscale 3D phononic crystal structures.

## Methods

Fig. 1 shows the structures of the isotopic nanoscale 3D phononic crystal, where [28]Si and [M]Si atoms are assembled periodically in three spatial directions. Fig. 1(a)-(d) shows the structures of 3D phononic crystals with different period lengths, corresponding to 1.09 nm, 2.17 nm, 3.26 nm and 6.52 nm, respectively. The volume of simulation cell is $12 \times 12 \times 12$ unit$^3$ (1unit is 0.543 nm), which has 13,824 atoms. The Green-Kubo method, equilibrium molecular dynamics (MD), are employed in calculating the thermal conductivities of 3D phononic crystal at 1000 K. MD simulation is a popular method in calculating thermal conductivity at high temperature.[20-22] In the following study, we focus on the thermal conductivity of 3D phononic crystal at 1000 K, which is larger than the Debye temperature, $T_D$, of Si (~658 K).[23]

In simulations, the periodic boundary condition is applied in all three directions. To derive the force term, we use Stillinger-Weber (SW) potential for Si,[24] which includes both two-body and three-body potential terms. The SW potential has been used widely to study the thermal properties of Si bulk material[15,21,25] for its accurate fit for experimental results on the thermal expansion coefficients. The heat current is defined as [20]

$$\vec{J}_l(t) = \sum_i \vec{v}_i \varepsilon_i + \frac{1}{2} \sum_{ij} \sum_{i \neq j} \vec{r}_{ij} \left( \vec{F}_{ij} \cdot \vec{v}_i \right) + \sum_{ijk} \vec{r}_{ij} \left( \vec{F}_j(ijk) \cdot \vec{v}_j \right) \tag{1}$$

where $\vec{F}_{ij}$ and $\vec{F}_{ijk}$ denote the two-body and three-body force, respectively. Thermal conductivity is calculated from the Green-Kubo formula [26]

$$\kappa = \frac{1}{3k_B T^2 V} \int_0^\infty < \vec{J}(\tau) \cdot \vec{J}(0) > d\tau \tag{2}$$

where $\kappa$ is thermal conductivity, $k_B$ is the Boltzmann constant, V is the system volume, T is the temperature, and the angular bracket denotes an ensemble average.

Generally, the temperature in MD simulation, $T_{MD}$, is calculated from the kinetic energy of atoms according to the Boltzmann distribution:

$$\langle E \rangle = \sum_1^N \frac{1}{2} m v_i^2 = \frac{3}{2} N k_B T_{MD} \tag{3}$$

Where $\langle E \rangle$ is the mean kinetic energy, $v_i$ is the velocity, m is the atomic mass, N is the number of particles in the system, and $k_B$ is the Boltzmann constant. This equation is valid at high temperature ($T \gg T_D$, $T_D$ is the Debye temperature).

Numerically, velocity Verlet algorithm is employed to integrate equations of motion, and each MD step is set as 1.0 fs. Firstly, canonical ensemble MD with langevin heat reservoir runs for $2^{20}$ steps to equilibrate the whole system at 1000 K. Then, microcanonical ensemble (NVE) MD runs for another $2^{25}$ steps (33.5 ns). Meanwhile, heat current is recorded at each step. At the end, the thermal conductivity is calculated by Eq. (2). In the calculation of thermal conductivity, the integration is from zero to a cut-off time which is determined by "first avalanche" method.[27] The final result is averaged over sixteen realizations with different initial conditions to satisfy ergodicity.

## Results and discussions

There is finite size effect in calculating thermal conductivity when the simulation cell is not big enough.[20,28] As shown in Fig. 2, we calculated the values of thermal conductivity of isotopic nanoscale 3D phononic crystal with different size of simulation cell, where the period length is 2 units and the mass ratio is 2. The calculated value of thermal conductivity converges when the side length of cubic simulation cell is larger than 10 units. To overcome the finite size effect on the calculated thermal conductivity, we use the side length of simulation cell as 12 units in the following simulations.

Fig. 3(a) shows the thermal conductivity of 3D phononic crystal with different period length. For comparison, we also calculate thermal conductivity of pure $^{28}$Si as $50 \pm 2$ W/m-K (the dash dot line in Fig. 3(a)), which is comparable to Schelling et al.'s results of MD simulation, 61 W/m-K.[20] However, MD results can not exactly coincide with the experimental value of $^{28}$Si at 1000 K [29], around 30 W/m-K, because of the inaccuracies of semi-empirical potentials and the impurity of the sample in measurements. This non-coincidence has little effect on the comparing MD results calculated using same potential parameters.

As shown in Fig. 3(a), the thermal conductivity rapidly decreases as the period length increases. The smallest value of thermal conductivity is 2.14 W/m-K, which is only 4.3% of pure $^{28}$Si calculated by EMD method. As shown in Fig.1 of Ref.[13], increasing period length may increase the amount of band folding and decrease the average velocity in the superlattice, resulting in a decrease of thermal conductivity. The phonon mean free path of Si is around 60 nm

*Reduction of Thermal Conductivity by Nanoscale 3D Phononic Crystal*

at 1000 K,[20] which is much longer than the period length in our simulation. The tendency of thermal conductivity in Fig. 3(a) is consistent with Simkin and Manhan's results when the phonon mean free path is larger than period length.[13]

Another way to modulate the phonon transport in the 3D phononic crystal is to vary the mass; because the mass of impurity atoms could perturb the phonon density of state and phonon dispersion curves which can affect the group velocities. However, our results indicate thermal conductivity rapidly decreases as the mass ratio increases from 1 to 6 (Fig. 3(b)). The smallest value of thermal conductivity is 0.54 W/m-K, which is only 1.1% of pure $^{28}$Si calculated by EMD method. The heaviest Si isotope atoms produced is $^{43}$Si.[30] Artificial Si isotopic atoms are used here to explore the mass influence on thermal transport and show the trend of large mass effects. $^{M}$Si atoms can be replaced by other heavier atoms, such as $^{54}$Fe.[31] When there are other kind atoms, the system is more complicated. That is, the mass is not the only factor involved. The bond strength and lattice relaxations must play a role, which is not studied in this letter.

In a $^{28}$Si$_{182}$ $^{M}$Si$_{10}$ quasi-1D supercell with 10 $^{M}$Si atoms (5%) randomly distributed, Gibbons and Estreicher[31] found the thermal conductivity decreased first and reached a minimum when the mass ratio was "two", and then the thermal conductivity increased as the increase of M. However, they stated that they cannot comment about the reasons for this minimum and do not know if the factor "two" remains valid for concentrations other than 5%. Different from randomly distributed, $^{M}$Si in 3D phononic crystal is periodic distributed in $^{28}$Si. The concentration of $^{M}$Si (50%) is much bigger than 5%. Our results show that the thermal conductivity of 3D phononic crystal decreases monotonously with increase of M. This is coincidence of the monotonous decrease of group velocities (Fig. 4(b)).

As shown above, changing the mass of impurity atoms and the period lengths are two effective ways in modulating the thermal conductivity. To find the mechanism in the decrease of thermal conductivity of 3D phononic crystal, the phonon dispersion curves are calculated through classical lattice dynamics. We calculated the dispersion curves by general utility lattice program (GULP),[32] and Stillinger-Weber potential [24] which is the same atom interaction as in our MD simulation. Fig. 4(a) shows acoustic branches and partial optical branches of the dispersion curves of the 3D phononic crystal with two different period lengths, 2 and 4 units. The Brillouin zone with 2 units in period length is double the size of that with 4 units. To compare two dispersion curves, the dispersion curves with 2 units in period length are folded in Γ to R direction to keep the same size of Brillouin zone as the one with 4 units. As the optical phonons contribute less to the thermal conductivity due to the lower group velocities, we focused on the acoustic phonons. It is clearly shown in Fig. 4(a) close to R point that the group velocities decrease as the period length increase, which causes the reduction of the thermal conductivity (shown in Fig. 3(a)). Fig. 4(b) shows acoustic branches of the 3D phononic crystal with different mass ratios. The dispersion curves are affected by the mass of impurity atoms and the group velocities decrease as the mass of impurity atoms increase, which contributes to the decrease of thermal conductivity (shown in Fig. 3(b)).

To understand more about the physical mechanism of thermal conductivity reduction, we carry out a vibration eigen-mode analysis on 3D phononic crystals. The mode localization can be qualitatively characterized by the normalized inverse participation ratio (NIPR).[33] The NIPR for phonon mode k is defined through the normalized eigenvector $u_k$

$$P_k^{-1} = N \cdot \sum_{i=1}^{N} \left( \sum_{\alpha=1}^{3} u_{i\alpha,k}^2 \right) \qquad (4)$$

where $N$ is the total number of atoms. When there are less atoms participating in the motion, the phonon mode has a large NIPR value. For example, NIPR is $N$ when there is only one atom vibrates in the localized mode. When all atoms participate in the motion, NIPR is calculated out as 1. That is, the larger of the value of NIPR the more localized of a phonon mode.

Fig. 5(a) shows the NIPR spectra of 3D phononic crystal with two different period lengths. Obviously, the NIPR for 3D phononic crystal with 4 units in period length has larger values than that with 2 units in period length. That is, there are more modes localized in 3D phononic crystal with 4 units in period length, which also leads to a reduction of its thermal conductivity. Fig. 5(b) shows the NIPR spectra of 3D phononic crystal with different mass ratio R. The values of NIPR for R=1.5, R=2 and R=2.5 are larger than pure Si (R=1), which show that the isotopic atoms could cause more localizations. NIPR values of 3D phononic crystals with different mass ratio R are close with each other. In Fig. 5(b), band gaps appear in the high frequency part (>12 THz) of spectra for R=2 and R=2.5, and the band gaps for R=2.5 are wider than those for R=2. The phonons with frequency in the range of band gaps cannot exist in the 3D phononic crystal.

## Conclusion

In this letter, using the Green-Kubo method, we have calculated thermal conductivities of isotopic nanoscale 3D phononic crystals where $^{28}Si$ and $^{M}Si$ atoms are assembled periodically in the three directions. Results show that the thermal conductivity decreases as the increasing of period length from 1 nm to 6 nm. The thermal conductivity of structure with 6nm period length is 2.14 W/m-K at 1000 K, which is only 4.3% of pure $^{28}Si$. Moreover, the thermal conductivity rapidly decreases as the mass ratio increases. The phonon localizations and bandgaps at high frequency in the 3D phononic crystal are shown clearly in the spectra of the normalized inverse participation ratio. The appearance of band gaps blocks a range of frequency of phonon modes and flattens phonon dispersion curves. The phonon dispersion curves show the phonon group velocities decrease in 3D phononic crystal. In a word, the decrease of thermal conductivity in 3D phononic crystal is attributed to both the decrease of group velocities and the localization.

If the bandgaps of the 3D phononic crystals happen at low frequencies, there will be greater reduction of the thermal conductivity because thermal conductivity is mainly contributed from acoustic phonons. Hence, manipulating band gaps to low frequencies in nanoscale 3D phononic crystal is a challenging work which is worthy of effort in the future. Although the isotopic atoms may have less effect on electric properties, it needs to find the effect on power factor by isotopic nanoscale 3D phononic crystal, when obtaining the exact ZT value of 3D phononic crystal. There are advances in obtaining the nanoscale 2D phononic crystal. However, the limitation in fabricating nanoscale 3D phononic crystal is still challenging nowadays.

Acknowledgements

This work was supported in part by a grant from the Asian Office of Aerospace R&D of the US Air Force (AOARD-114018) (LY and BL), and in part by a startup fund from Tongji University (NY). NY and LY wishes to acknowledge useful discussions with Jin-Wu Jiang (Bauhaus-University Weimar), Jie Chen (NUS), Sha Liu (NUS), and Lifa Zhang (NUS).

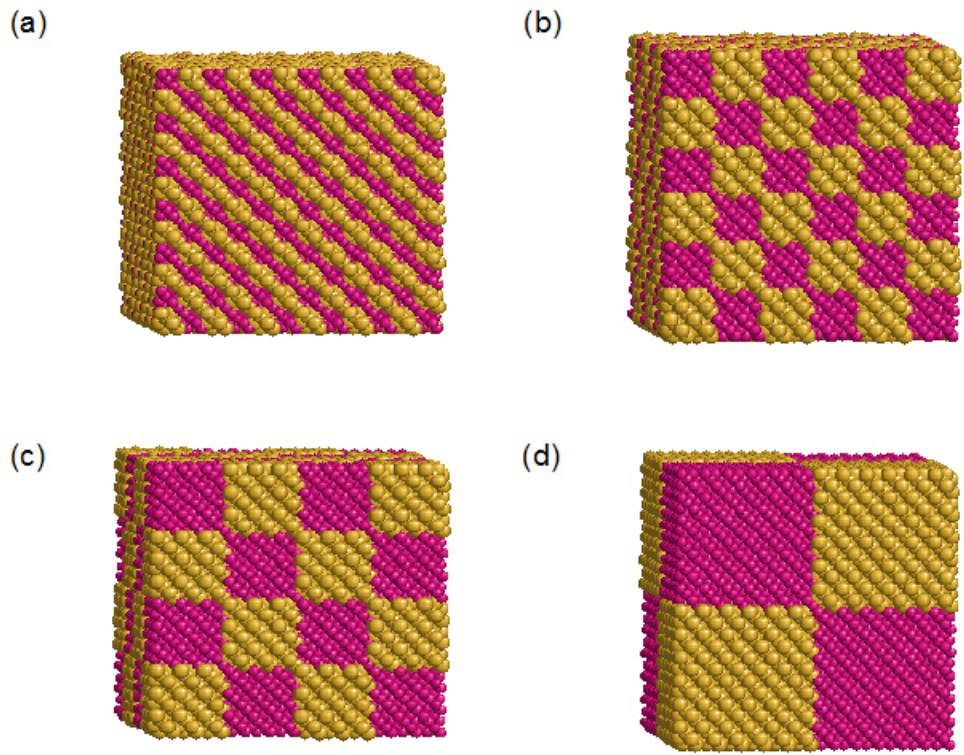

FIG. 1. (Color online) The structures of the isotopic nanoscale 3D phononic crystals, three dimensional periodic arrangements of $^{28}$Si (pink) and $^M$Si (yellow) atoms. From (a) to (d), the period lengths of those 3D phononic crystals are 2, 4, 6 and 12 units, respectively. The lattice constant is 0.543 nm, that is, 1 unit represents 0.543 nm.

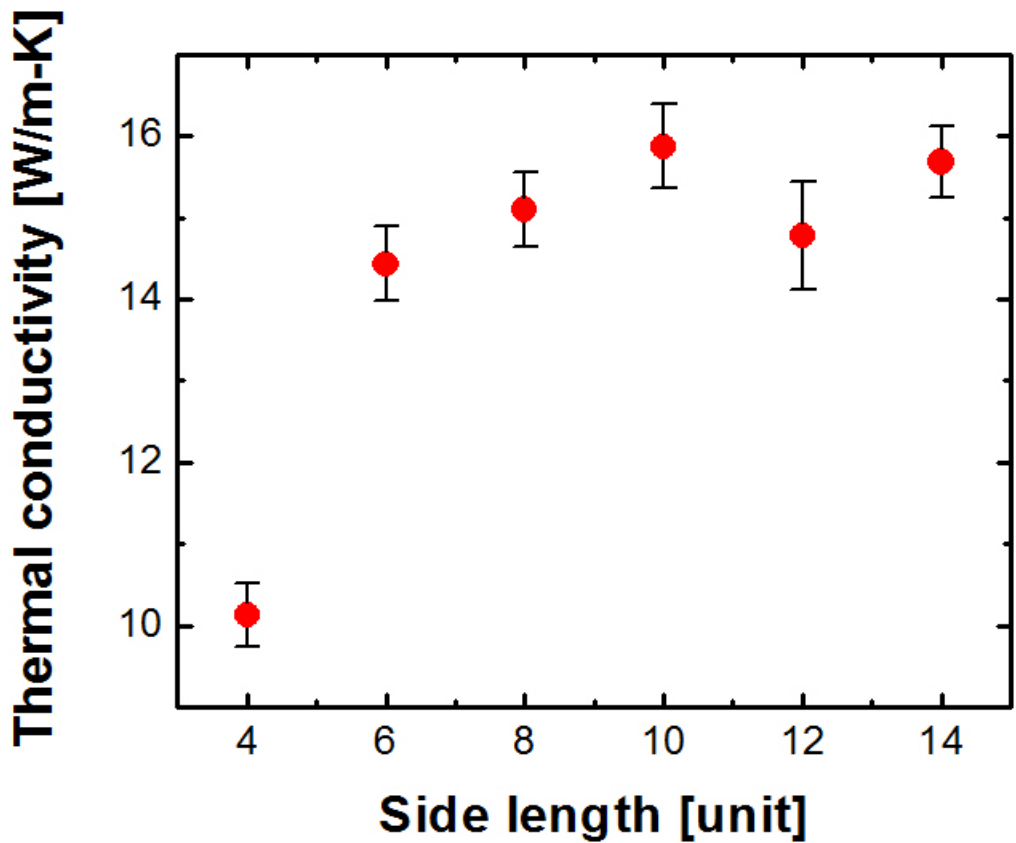

FIG. 2. (Color online) Thermal conductivity versus the side length of simulation cell of isotopic nanoscale 3D phononic crystal of Si. The mass ratio is 2 and the period length is 2 units. The error bars are calculated from 16 simulations with different initial conditions. All values are calculated at 1000 K.

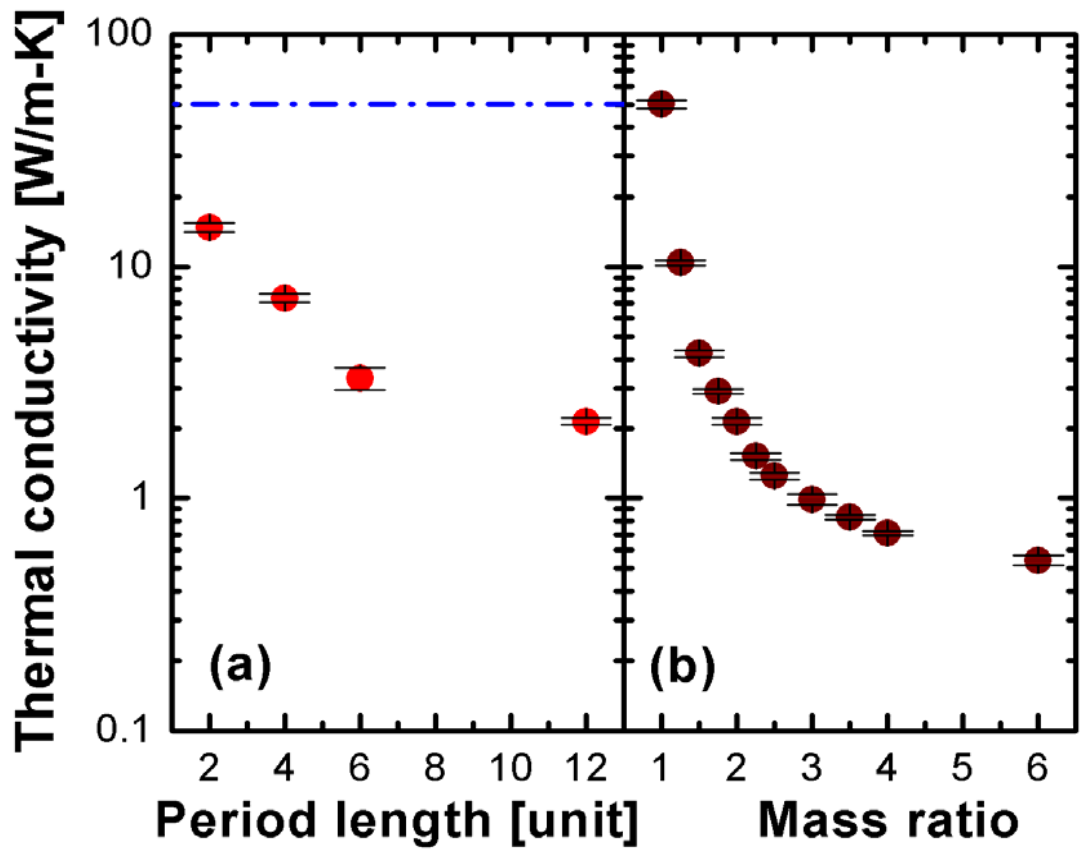

FIG. 3. (Color online) (a) Thermal conductivity versus the period length of isotopic nanoscale 3D phononic crystal of Si. The mass ratio is 2. The dash dot line corresponds to the molecular dynamic result of thermal conductivity of pure $^{28}$Si. (b) Thermal conductivity versus the mass ratio of isotopic nanoscale 3D phononic crystals of Si. The period length is 12 units. All values are calculated at 1000 K.

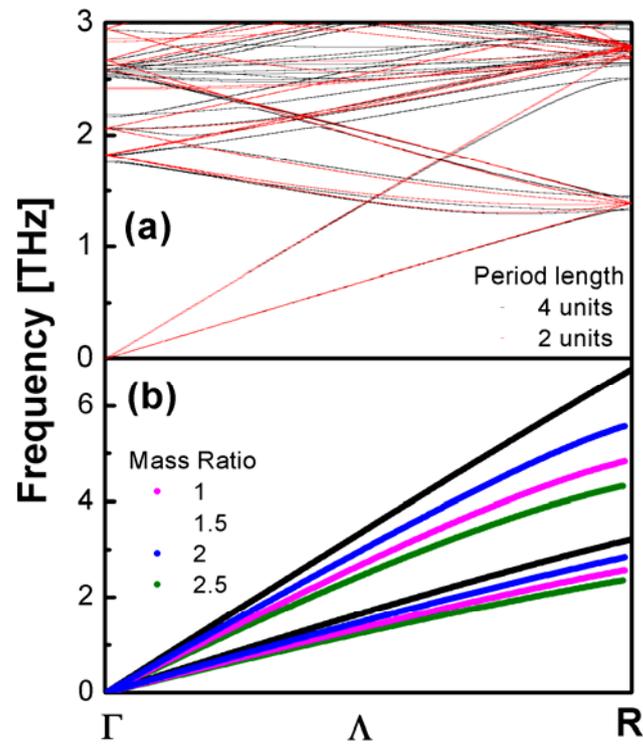

FIG. 4. (Color online) (a) Acoustic and partial optical branches along the [1, 1, 1] direction. The mass ratio of the isotopic nanoscale 3D phononic crystal of Si is 2. Different colors present dispersions curves of 3D phononic crystal with different period length. It is clearly shown close to R point that the decrease of the group velocities as the period length increase, which causes the reduction of the thermal conductivity. (b) Acoustic branches along the [1, 1, 1] direction. The mass ratio of 3D phononic crystals changes from 1 to 2.5. Different color is referred to different mass ratio. The period lengths of 3D phononic crystals are kept same, 2 units. The dispersion curves are affected by the mass of impurity atoms and the group velocities decrease as the mass of impurity atoms increase, which contributes to the decrease of thermal conductivity.

*Reduction of Thermal Conductivity by Nanoscale 3D Phononic Crystal*

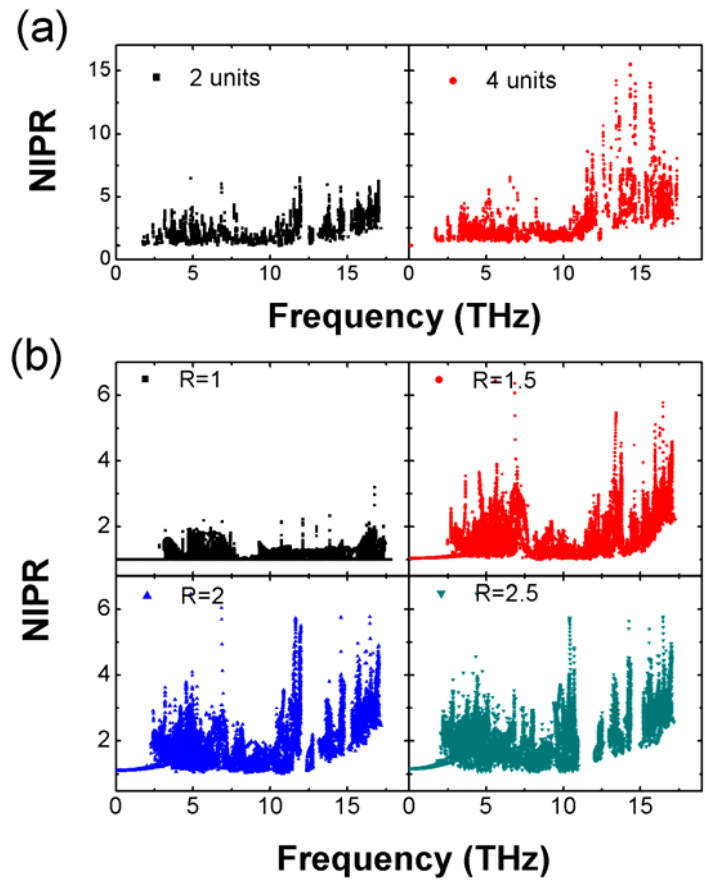

FIG. 5. (Color online) The normalized verses participation ratio (NIPR) spectra. NIPR is calculated based on Eq. (4). The larger of the value of NIPR the more localized of a phonon mode. (a) NIPR spectra of 3D phononic crystals with different period length. The left and right panels are corresponding to 3D phononic crystals with 2 unit and 4 unit period length, respectively. The mass ratios are same as 2. (b) NIPR spectra of 3D phononic crystals with different mass ratio, R. The period lengths are same as 2 units. The upper left panel (R=1) corresponds to pure Si.